\documentclass[reprint,aps,prl]{revtex4-2}

\usepackage{graphicx}% Include figure files
\usepackage{dcolumn}% Align table columns on decimal point
\usepackage{bm}% bold math
\usepackage{amsmath}
\usepackage{amssymb}
\usepackage{subcaption}
\usepackage{bbold}
\usepackage{orcidlink}
\usepackage{bbding}
\usepackage{slashed}
\usepackage{float}
\usepackage{verbatim}
\usepackage{hyperref}
\usepackage{multirow}
\begin{document}

%\preprint{APS/123-QED}

\title{CP violation angles from H$\to\tau\tau$ decays at FCC-ee}

\author{S. Giappichini\orcidlink{0009-0002-7694-641X}}
\author{M. Klute\orcidlink{0000-0002-0869-5631}}
\author{M. Presilla\orcidlink{0000-0003-2808-7315}}
\affiliation{Institute for Experimental Particle Physics (ETP), Karlsruhe Institute of Technology (KIT), Wolfgang-Gaede-Straße 1, Karlsruhe, D-76131, Germany}

\date{\today}

\begin{abstract}
\noindent Measuring charge-parity (CP) violation in Higgs-fermion interactions is a key target of future precision Higgs programs. The decay $H\to\tau\tau$ is particularly sensitive to the CP structure of the Higgs Yukawa coupling via $\tau$ spin correlation, while the clean environment of $e^+e^-$ collisions at the Future Circular Collider (FCC) enables accurate reconstruction of CP-sensitive observables. In this letter, we study the sensitivity of FCC-ee to the Higgs CP state using $H\to\tau\tau$ events produced in associated $ZH$ production at $\sqrt{s}=240$ GeV. In the anomalous-coupling parametrization, we project a precision of $\Delta\phi_{\tau\tau}=\pm 2.5^\circ$ at 68\% CL, with the dominant contribution arising from one-prong hadronic $\tau$ decays. We further interpret the analysis in the Effective Field Theory framework, deriving expected limits on the CP-odd operators, and compare them with electron and magnetic dipole moment measurements and the anomalous-coupling approach.
\end{abstract}

\maketitle
%\textit{Introduction}. 
The exploration of the Higgs boson sector is the current top priority for the high-energy particle physics community \cite{Altmann:2025feg,deBlas:2025gyz} to address major fundamental questions in the field. 
Consequently, several projects targeting the production of Higgs bosons from $e^+e^-$ collisions have emerged, taking advantage of the clean environment and new technologies to achieve higher luminosities. The Future Circular Collider (FCC-ee) \cite{fcclepton,FCC:2025uan} has been individuated as the next flagship collider at CERN \cite{Arduini:2947728}, with operations oreseen to begin in the second half of the century. The FCC-ee program's core objective is to inclusively measure Higgs production with a dedicated run at $\sqrt{s}=$ 240 GeV with $\mathcal{L}_{\mathrm{int}} = 10.8\; \mathrm{ab}^{-1}$, where approximately two million Higgs bosons will be produced in association with Z bosons (ZH) \cite{fccphysics,FCC:2025lpp}. This will enable model-independent precision measurements of the Higgs boson's properties \cite{Selvaggi:2025kmd}. One such measure is the charge-parity violation (CPV) in Higgs interactions. Although the LHC has confirmed that the spin-parity quantum numbers of the Higgs boson align with the SM prediction of $J^{PC} = 0^{++}$ \cite{higgsparity}, the possibility of small anomalous CPV contributions remains \cite{ATLAS:2017azn,ATLAS:2018hxb,ATLAS:2023mqy,CMS:2016tad,CMS:2019ekd,CMS:2019jdw}, which would indicate physics beyond the Standard Model (BSM) and lead to the understanding of the observed baryon asymmetry of the Universe \cite{Wagner:2023vqw,Guo:2016ixx,King:2015oxa}. Thus, one of the goals for future colliders is to test CPV with absolute precision greater than $10^{-2}$ in fermion vertices and $10^{-6}$ in vector boson channels \cite{snowmass_Higgs}.

This letter explores the sensitivity of the FCC-ee to the CP property of the H$\to\tau\tau$ Yukawa interactions using two different approaches. The first approach, known as the anomalous coupling method in literature, introduces CPV via an effective Lagrangian that directly modifies the H$\to\tau \tau$ coupling \cite{Harnik:2013aja,Brod:2013cka}. The second approach relies on the Standard Model Effective Field Theory (SMEFT)~\cite{BuchmuellerWyler, warsaw}, where higher-dimensional operators in the SM Lagrangian reflect the low-energy effects of heavy new-physics degrees of freedom beyond the collider reach in a model-agnostic way. This enables the current experimental measurement to constrain a broader spectrum of BSM scenarios, allowing global fits to simultaneously probe both CP-even and CP-odd contributions.

CP is a composite quantity, arising from CP-even and CP-odd contributions, and is not directly observable. Instead, it can be probed via spin correlation in Higgs decays mediated by fermionic vertices at tree level, while interactions with vector bosons are loop suppressed \cite{Ge:2020mcl,Fuchs:2020uoc,DeVries:2018aul}.
In this context, the H$\to\tau\tau$ vertex provides an excellent probe of CP interactions, also benefiting from the largest leptonic branching fraction. 
The CP nature of the Higgs boson is reflected in the spin correlation of the $\tau$ leptons produced in its decay, with transverse polarization correlations providing sensitivity to CP mixing \cite{Desch:2003rw}, while the longitudinal polarization components probe the spin of the parent particle.
The $\tau$ polarization information can be inferred from its decay products \cite{Tsai:1971vv}, allowing the construction of CP-sensitive observables based on the kinematics of the reconstructed final state \cite{Grzadkowski:1995rx}.

Current measurements at the LHC place relatively loose constraints on the CP-mixing angle in the anomalous coupling approach, yielding values of $-1^\circ \pm 19^\circ$ from CMS \cite{CMS:2021sdq} and $9^\circ \pm 16^\circ$ from ATLAS \cite{ATLAS:2022akr}. Both results disfavor a purely CP-odd Higgs boson at a significance exceeding three standard deviations.
Prospects for the HL-LHC, assuming an integrated luminosity of $\mathcal{L}_{\mathrm{int}} = 3\;\mathrm{ab}^{-1}$ at $\sqrt{s} = 14$ TeV, indicate that the precision on the CP-mixing angle could improve to approximately $\pm 8^\circ$ \cite{Harnik:2013aja,Esmail:2024jdg,deBlas:2025gyz,ATLAS:2019gtt}. By comparison, studies for the International Linear Collider (ILC), an $e^+e^-$ collider with polarized beams operating at $\sqrt{s} = 250$ GeV, project a precision of $\pm 4.3^\circ$ using only one-prong hadronic $\tau$ decays \cite{Jeans:2018anq}.

The FCC-ee offers excellent sensitivity to Higgs couplings through the $ZH$ production mechanism. At $\sqrt{s}=240$ GeV, measurements of $H \to \tau\tau$ are expected to achieve a signal strength precision of approximately $0.57\%$ for $\sigma_{ZH}\times\mathcal{B}(H\to\tau\tau)$, corresponding to a relative precision of $0.47\%$ on the Higgs–tau coupling modifier $\kappa_\tau$ \cite{Giappichini:2026vlg}. While such rate measurements are not directly sensitive to CP mixing, they provide stringent constraints on the allowed parameter space for Higgs CP violating interactions.

\begin{figure}[htb]
    \centering
    \includegraphics[width=0.7\columnwidth]{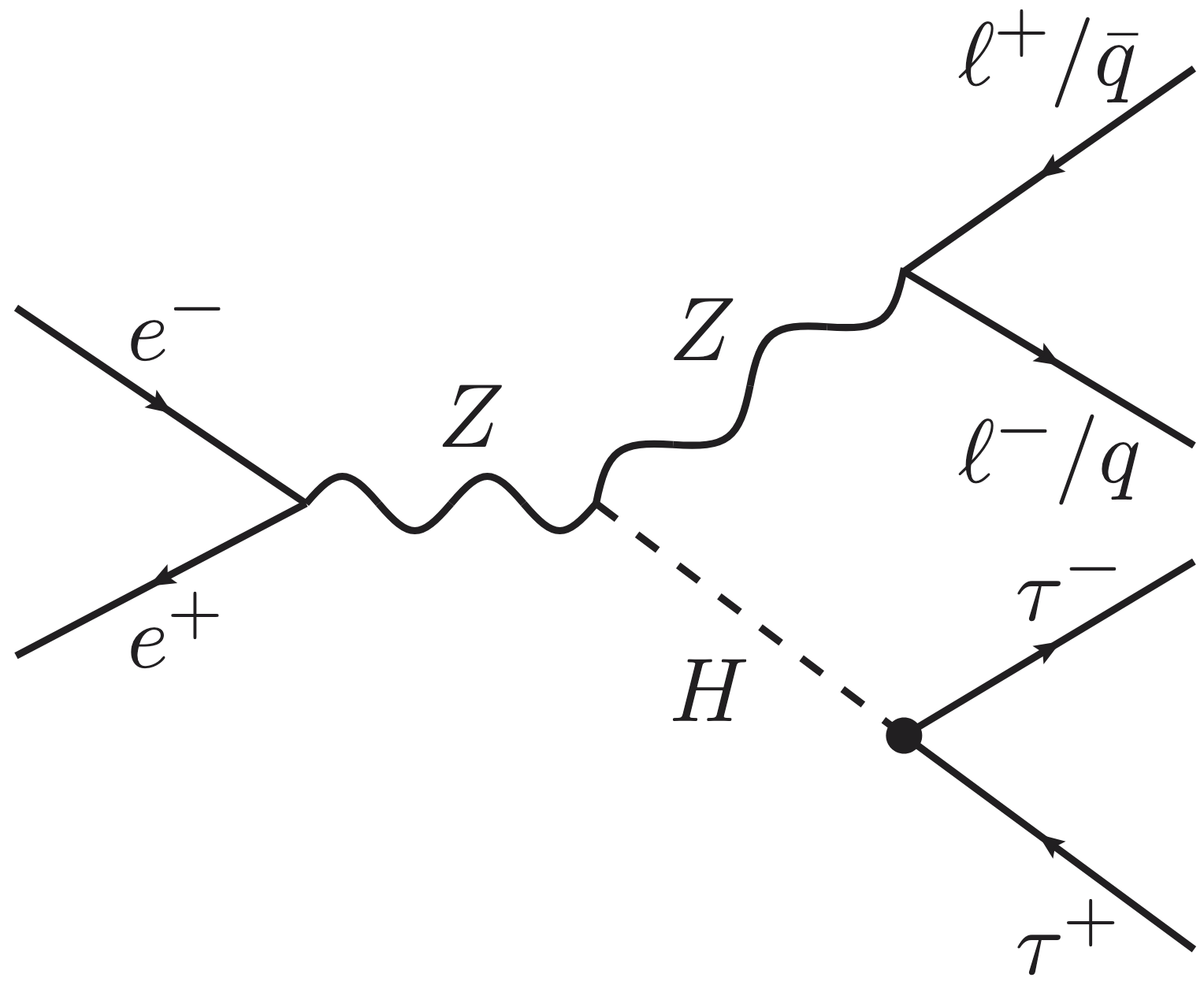}
    \caption{Feynman diagram for ZH events in $e^+e^-$ collisions, where the Z boson decays to lepton (electrons or muons considered in this study) or quark pairs and the Higgs decays to a $\tau$ pair. An additional source of CPV (represented by a black dot) could modify the $H \tau \tau$ interaction.}
    \label{fig:diagram}
\end{figure}

\bigskip
\textit{The anomalous coupling formalism}. CPV in the H$\to\tau\tau$ vertex is introduced through an anomalous Yukawa coupling that models both CP-even and CP-odd interactions \cite{Harnik:2013aja,Brod:2013cka,CMS:2014nkk,Askew:2015mda}. The corresponding Lagrangian is written as 
\begin{equation}
    \mathcal{L}_{H\tau\tau}=-\frac{m_\tau}{v}\kappa_\tau\bar{\tau}(\cos\phi_{\tau\tau}+i\gamma_5\sin\phi_{\tau\tau})\tau H,
\end{equation}
where $m_\tau$ is the mass of the $\tau$ lepton, $v$ is the vacuum expectation value, and
\begin{equation}
    \phi_{\tau\tau}=\arctan\left(\frac{\tilde{\kappa}_\tau}{\kappa_\tau}\right)
\end{equation}
is the CP mixing angle used to parametrize the mixing between the reduced CP-even Yukawa coupling 
\begin{equation}
    \kappa_\tau=\frac{v}{m_\tau}y_\tau,
\end{equation}
with $y_\tau$ being the SM Yukawa coupling, and its CP-odd counterpart $\tilde{\kappa}_\tau$. The pure CP-even Higgs boson corresponds to $\phi_{\tau\tau}=0^\circ$, while a pure CP-odd Higgs is given by $\phi_{\tau\tau}=90^\circ$, with intermediate values corresponding to mixed states for $\kappa_\tau=1$. The overall sign of the Lagrangian is physically irrelevant, making it insensitive to a global minus factor.

\bigskip
\textit{The Standard Model Effective Field Theory approach.} The EFT Lagrangian can be expressed as ~\cite{BuchmuellerWyler, warsaw}:
\begin{equation} \label{SMEFTlagrangian}
    \mathcal{L}_{\mathrm{EFT}}=\mathcal{L}_{\mathrm{SM}}+\sum_d \sum_i \frac{c_i^{(d)}}{\Lambda^{d-4}} \mathcal{O}_i^{(d)}, 
\end{equation}
with $d>4$. The parameters $c_i^{(d)}$ are the Wilson coefficients associated with the corresponding operators $\mathcal{O}_i^{(d)}$ of dimension $d$. The parameter $\Lambda$ sets the scale of the BSM physics. 

In scenarios that conserve both baryon and lepton number, only even-dimensional operators are retained. Truncating the expansion at dimension six, the leading order beyond the SM, the cross-section for a given process can be expressed as
\begin{gather}
    \sigma_{\mathrm{SM}+\mathrm{EFT}} \propto\left|\mathcal{M}_{\mathrm{SM}}+\sum_i \frac{c_i}{\Lambda^{2}} \mathcal{M}_{i} \right|^2 = \\ \notag
    |\mathcal{M}_{\mathrm{SM}}|^2 + 2 \sum_{i} \frac{c_i}{\Lambda^2} \mathcal{R}e(\mathcal{M}_{\mathrm{SM}}^* \mathcal{M}_i) + \sum_{i,j} \frac{c_i c_j}{\Lambda^4} (\mathcal{M}_i^* \mathcal{M}_j).
\end{gather}
The interference term, proportional to 
$\Lambda^{-2}$, represents the leading BSM correction to the SM cross-section, while the term of order $\Lambda^{-4}$ corresponds to contributions quadratic in the dimension-six amplitudes. Contributions from higher-dimensional terms are assumed to be suppressed by additional powers of $\Lambda$ and are therefore neglected.

The interference term $ 2 \sum_{i} \frac{c_i}{\Lambda^2} \mathcal{R}e (\mathcal{M}_{\mathrm{SM}}^* \mathcal{M}_i)$ is CP-odd for CP-odd interactions and thus contributes to asymmetries in such observables. At the same time, it vanishes in any CP-even quantity integrated over the full phase space. Conversely, quadratic terms in the CP-odd amplitudes are CP-even, leading only to modifications of total cross-sections. Hence, probing CP-odd observables is essential to extract genuine constraints on CPV effects in gauge-Higgs interactions.

In the Warsaw basis~\cite{warsaw}, CPV in the SMEFT can originate either from operators that are odd under CP with real Wilson coefficients, or from operators that are not self-conjugate and therefore allow complex Wilson coefficients, with CPV encoded in their imaginary parts.
In this work, we study CPV in electroweak interactions by considering three Hermitian dimension-six CP-odd operators coupling the Higgs doublet to electroweak gauge bosons, together with a non-self-conjugate operator governing the Higgs–tau interaction.

The tau-Yukawa operator can be generally written as
\begin{equation*}
\mathcal{O}_{eH} = (H^\dagger H) (\bar{L}_L \tilde{H} e_R) + \text{h.c.}
\end{equation*}
In this definition, $L_L$ represents the left-handed lepton doublet, $e_R$ is the right-handed lepton, $H$ is the Higgs doublet, and $\tilde{H}$ is the conjugate Higgs doublet. The imaginary part of this effective Yukawa coupling, $\mathcal{I}m (\mathcal{O}_{eH})$, induces CPV in the Higgs decay.

The CP-odd HVV operators, which could provide an additional source of CPV in the process, are
\[
\begin{aligned}
\mathcal{O}_{H \widetilde{W}} & =\left(H^{\dagger} H\right) \widetilde{W}_{\mu \nu}^I W^{I \mu \nu} \\
\mathcal{O}_{H \widetilde{W} B} & =\left(H^{\dagger} \tau^I H\right) \widetilde{W}_{\mu \nu}^I B^{\mu \nu} \\
\mathcal{O}_{H \widetilde{B}} & =\left(H^{\dagger} H\right) \widetilde{B}_{\mu \nu} B^{\mu \nu}
\end{aligned}
\]
where $H$ is the Higgs doublet, $W_{\mu \nu}^I$ and $B_{\mu \nu}$ are the $S U(2)_L$ and $U(1)_Y$ field strengths, $\tau^I$ are the Pauli matrices, and $\tilde{X}_{\mu \nu} \equiv \frac{1}{2} \varepsilon_{\mu \nu \rho \sigma} X^{\rho \sigma}$ denotes the dual tensor. 

For completeness, the corresponding CP-even operators $\mathcal{O}_{HW}$, $\mathcal{O}_{HB}$, $\mathcal{O}_{HWB}$, $\mathcal{R}e(\mathcal{O}_{eH})$ are also included in the analysis.
 
\bigskip
\textit{Event simulation and selection.} This work analyzes Monte Carlo–simulated events generated at leading order.
For the signal processes in the anomalous coupling approach, three angular hypotheses ($\phi_{\tau\tau}=0^\circ, \;45^\circ \text{ and }90^\circ$) are simulated with the event generator Pythia8 \cite{pythia}. 
By contrast, the EFT signal processes are generated using the SMEFTsim 3.0~\cite{Brivio:2017btx,Brivio:2020onw} model with the TopU3L flavor scheme~\cite{Brivio:2021yjb} in Madgraph 3.5.1 \cite{madgraph}, in conjunction with the TauDecay package \cite{Hagiwara:2012vz} to model the helicity amplitudes for each $\tau$ decay mode.

The dominant background processes, $e^+e^- \to ZZ$, $e^+e^- \to WW$, and $e^+e^- \to q\bar q$ via $\gamma/Z$ exchange, are generated using Pythia8 \cite{pythia}. Initial- and final-state radiation are included consistently for all simulated signal and background samples.
The remaining background sources considered include Higgs decays from ZH production (ZZ, WW, gg, $b\bar{b}$, $c\bar{c}$, and $s\bar{s}$), Drell-Yan to leptons, $\gamma\gamma\to \ell\ell$, $e\gamma \to eZ$ with Z decaying into electrons or muons, and $ee\to \nu\nu Z$. These are simulated with Whizard v. 3.0.3 \cite{whizard1,whizard2} in conjunction with Pythia6 \cite{pythia6} to hadronize the events. Finally, the fast-simulation package Delphes v. 3.5.1pre05 \cite{delphes} simulates the response of the IDEA detector concept~\cite{IDEAStudyGroup:2025gbt}, one of the main designs proposed for FCC-ee based on a large volume drift wire chamber and a dual-readout calorimeter. 

We separate the analysis into six different event categories, depending on the decay of the Z bosons and the two $\tau$ leptons. The reconstruction of $\tau$ leptons depends on their decay mode: leptonic $\tau$ decays ($\tau_\ell$) have been reconstructed starting from the isolated leptons in the event, while hadronic $\tau$ leptons ($\tau_h$) are built from jets. These are handled by the package FastJet \cite{fastjet}, using the $e^+e^-$ exclusive Durham algorithm with $n_{jets}$ equal to the expected number of jets in each topology. The selection of the final state objects differs in each signal category, as well as the specific particles being clustering candidates. On top of this, we adopt a kinematic selection on the reconstructed objects and event observables to reduce the background, while maintaining a good signal efficiency. The details are given in Table \ref{tab:cuts-ll-qq}. Two important discriminating quantities related to the Higgs boson are the invariant mass of the $\tau$ pair, calculated using the collinear approximation \cite{collinear}, and the mass of the system recoiling against the $\tau$ lepton pair, extracted from the momenta of the decay products of the Z boson and total energy-momentum conservation. 

\begin{table*}[ht]
\centering
\resizebox{\textwidth}{!}{
\begin{tabular}{|l|ccc|ccc|}
\hline
 & \multicolumn{3}{c|}{$Z\to\ell\ell$} & \multicolumn{3}{c|}{$Z\to qq$} \\
 x& H$\to\tau_\ell\tau_\ell$ & H$\to\tau_\ell\tau_h$ & H$\to\tau_h\tau_h$ & H$\to\tau_\ell\tau_\ell$ & H$\to\tau_\ell\tau_h$ & H$\to\tau_h\tau_h$ \\
\hline
Object selection &
\begin{tabular}{@{}c@{}}4 leptons, total\\ neutral charge,\\ no hadrons/clustering\end{tabular} &
\begin{tabular}{@{}c@{}}3 leptons, 2 with\\ opposite sign,\\ excluded from 1 jet\end{tabular} &
\begin{tabular}{@{}c@{}}2 leptons, same flavor,\\ opposite sign,\\ excluded from 2 jets\end{tabular} &
\begin{tabular}{@{}c@{}}2 isolated leptons\\ ($p>20$ GeV)\\ excluded from 2 jets\end{tabular} &
\begin{tabular}{@{}c@{}}1 isolated lepton\\ ($p>20$ GeV)\\ excluded from 3 jets\end{tabular} &
\begin{tabular}{@{}c@{}}4 jets from\\ all particles\end{tabular} \\
\hline
$100<M_{collinear}<150$ GeV & \checkmark & \checkmark & \checkmark & \checkmark & \checkmark & \checkmark \\
$115<M_{recoil}<160$ GeV & \checkmark & \checkmark & \checkmark & \checkmark & \checkmark & \checkmark \\
$\Delta R_{\tau\tau}>2$ rad & \checkmark & \checkmark & \checkmark & \checkmark & \checkmark & \checkmark \\
$\cos\theta_{\tau\tau}<-0.6$ & \checkmark & \checkmark & \checkmark & \checkmark & \checkmark & $-0.98<\cos\theta_{\tau\tau}<-0.6$ \\
$|\cos\theta_{miss}|<0.98$ & \checkmark & \checkmark & \checkmark & \checkmark & \checkmark & \checkmark \\
$80<M_Z<100$ GeV & \checkmark & \checkmark & \checkmark & \checkmark & \checkmark & \checkmark \\
$40<p_Z<55$ GeV & -- & -- & -- & \checkmark & \checkmark & \checkmark \\
$E_{miss}$ & $>20$ GeV & -- & -- & $>10$ GeV & $>10$ GeV & $>10$ GeV  \\
$E_H$ & $<110$ GeV & -- & -- & $<95$ GeV & -- & -- \\
\hline
\end{tabular}
}
\caption{Summary of object definition, jet clustering, and kinematic selection for each category. The first row describes the object selection and clustering; subsequent rows list the applied cuts. Leptons, $\ell$, are either electrons or muons, $q$ stands for all quarks except the top.}
\label{tab:cuts-ll-qq}
\end{table*}

The impact parameter vectors of charged tracks necessary for the rest of the study are computed by reconstructing the interaction point from the Z decay products' tracks, so there is no bias introduced by the presence of displaced tracks from the $\tau$ leptons. To differentiate the leptonic Z or $\tau_\ell$ decays, where there could be ambiguity, we first identify the opposite charge and same flavor pair with the invariant mass closest to the Z boson, then assign the rest to represent the $\tau_\ell$. 

Since reconstructing the CP angular variables, as will be seen later in more detail, requires the knowledge of the individual $\tau$ decay products, we employed an explicit $\tau_h$ reconstruction method~\cite{Giappichini:2026vlg}. The method directly classifies all jets containing electrons or muons as quark jets. The $\tau_h$ four-vector is reconstructed from constituents within $\Delta \theta < 0.2$ to the leading charged pion in the jet. Constraints requiring a net charge of $\pm 1$ and an invariant mass below 3~GeV are applied, such that the reconstructed $\tau_h$ generally constitutes only a fraction of the original jet. Information related to the charged prongs (one- or three-prong) and neutral hadrons is stored separately for later use.

\bigskip
\textit{Reconstruction of the CP sensitive angle}. 
The spin correlation of the $\tau$ leptons produced in Higgs decays encode information on the CP structure of the Higgs–tau Yukawa coupling. In particular, CP-sensitive observables can be constructed from the angle between the $\tau$ decay planes in the Higgs rest frame or from observables sensitive to $\tau$ polarization. Access to the full polarization information is straightforward if the $\tau$ four-momenta are fully reconstructed.
One example of this is presented in Ref. \cite{Jeans:2018anq}, which proposes to use the energy conservation of the events to reconstruct the full $\tau$ energy-momentum vectors \cite{Jeans:2015vaa} and applies it to the CP estimation. Unfortunately, this method can only be applied to $\tau_h$ decays.

We choose to use the methods developed in Ref. \cite{Berge:2008dr, Berge:2015nua, Bower:2002zx} to approximate the angle between the decay planes, given only the visible part of the $\tau$ decay, applying it to all decay modes. The charged prongs in one-prong decays (including leptons in leptonic decays), or the pion with the same sign as the $\tau$ identified as coming from the $\rho^0$ resonance by finding the pair that matches the invariant mass in three-prong decays, are used to define a zero-momentum frame (ZMF) that represents the Higgs rest frame. Then the angle between the planes in the ZMF is defined as
\begin{equation}
    \phi^{ZMF}=\arccos(\hat{\lambda}^{+ ZMF}_{\perp} \cdot \hat{\lambda}^{- ZMF}_{\perp}),
\end{equation}
where $\lambda^{\pm }$ are the impact parameter vectors if no neutral particle is present in the reconstructed $\tau$, the sum of the neutral particles if the decay is $\pi^\pm\pi^0\nu$ and $\pi^\pm\pi^0\pi^0\nu$ decays, or the pion with opposite charge to the reconstructed $\tau$ for the $\pi^\pm\pi^\pm\pi^\mp\nu$ decay. Then $\hat{\lambda}^{\pm ZMF}_{\perp}$ is the perpendicular component of the normalized vector with respect to the charged pion track in the ZMF. 

Additional discrimination is needed to avoid destructive interference between differently polarized states of the meson and to get the same periodicity of the angle in the range $[-\pi,\pi]$. The variable $y^\tau$ is defined as
\begin{equation}
    y^{\tau^\pm}=\frac{E_{\pi^\pm} - E_{\pi^0}}{E_{\pi^\pm} + E_{\pi^0}}, \qquad y^\tau=y^{\tau^+}\cdot y^{\tau^-},
\end{equation}
which is used in the three-prong decay by replacing $\pi^\pm$ with the prong used in the calculation of the ZMF, and the oppositely charged pion replaces $\pi^0$. Since $y^{\tau^\pm}=1$ in other cases, the first step of discrimination becomes
\begin{equation}
    \phi^{int}=\begin{cases}
        \phi^{ZMF} & \quad \text{if } y^\tau\geq 0 \\
        -\pi + \phi^{ZMF} & \quad \text{if } y^\tau<0
    \end{cases}\quad .
\end{equation}
The selection is inverted if only one of the two $\tau$ leptons decays leptonically, since the respective spectral functions have opposite signs from the ones for single-pion decays, which causes a phase flip between the corresponding $\phi_{int}$ distributions. Finally,  
\begin{equation}
    O^{ZMF}= \hat{q}^{-ZMF}\cdot (\hat{\lambda}^{+ZMF}_{\perp} \times \hat{\lambda}^{- ZMF}_{\perp}),    
\end{equation}
where $\hat{q}^{-ZMF}$ is the normalized four-momentum vector of the negative prong used for the ZMF calculation, is used to properly assign the periodicity of single-prong decays. The CP sensitive angle is expressed as
\begin{align}
    \phi_{CP}=\begin{cases}
        \phi^{int} & \quad \text{if } O^{ZMF}\geq 0 \\
        -\phi^{int} & \quad \text{if } O^{ZMF}<0
    \end{cases}\quad.
\end{align}

The dominant background process of this analysis is ZZ, where one of the bosons decays to two prompt $\tau$ leptons. The Z boson has a spin of 1, which means that the primary spin correlation involves the longitudinal polarizations of the $\tau$ particles, rather than their transverse ones, so the differential cross section does not depend on $\phi_{CP}$ \cite{Berge:2014sra}. The rest of the backgrounds, mostly coming from misidentified $\tau$, are also expected to be flat in the CP-sensitive variables, thus making the distinction from the signal very clear.

The most sensitive decay for this method is $\tau^\pm\to\pi^\pm\nu$. Since the pion has a spin of 0, the spin of the $\tau$ is transferred to the neutrino, resulting in a well-defined helicity state that anti-propagates the $\tau$ spin direction. The pion's momentum unambiguously identifies the neutrino's flight direction, thus encoding the $\tau$ spin in an experimentally accessible way. In contrast, other decays dilute this power, especially leptonic decays due to the presence of two neutrinos. Therefore, EFT results will be presented only for one-prong hadronic decay final states. 

\begin{figure}[htb]
    \centering
    \begin{subfigure}[h]{\columnwidth}
    \centering
        \includegraphics[width=0.7\columnwidth]{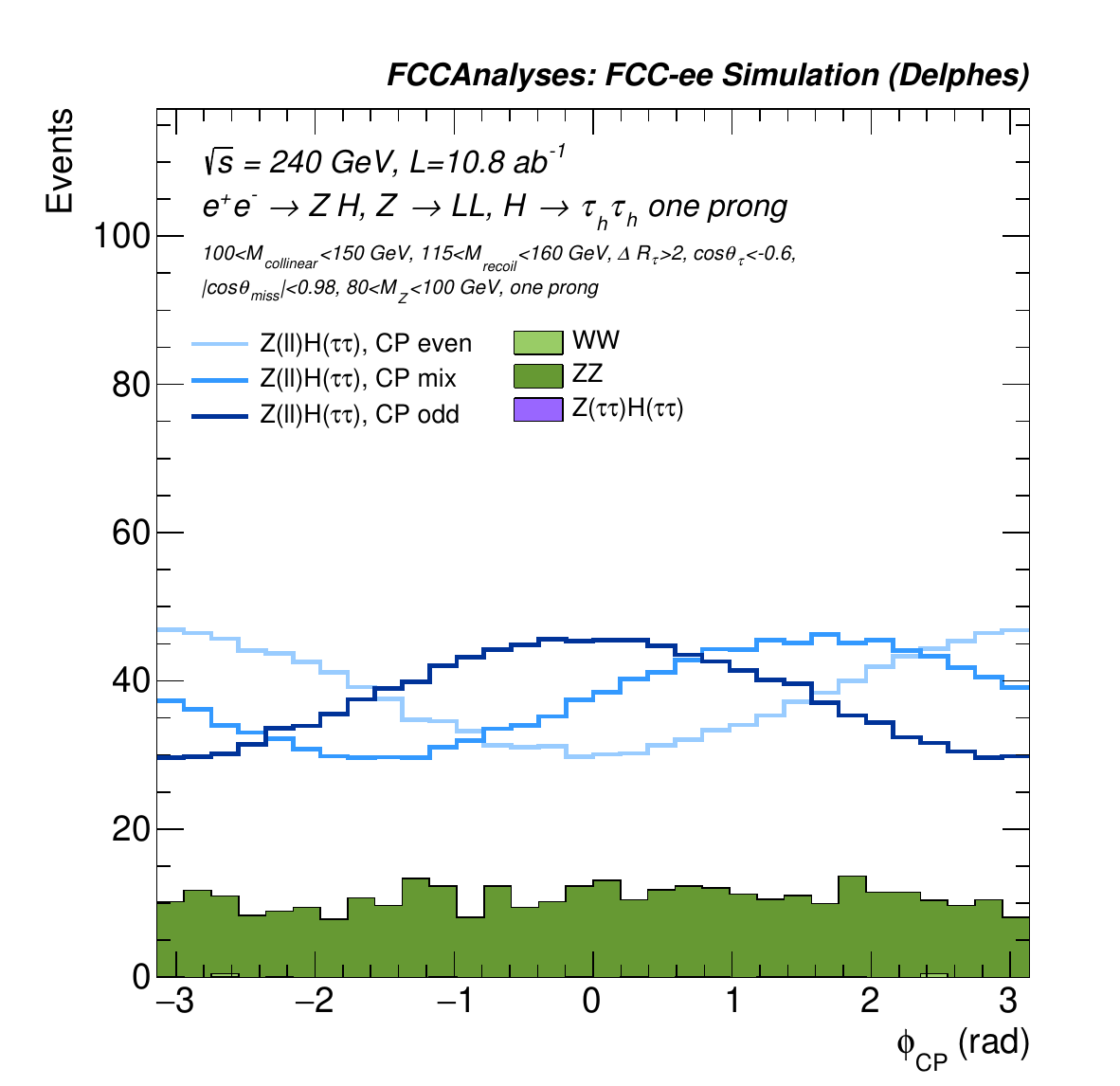}
    \end{subfigure}
    \begin{subfigure}[h]{\columnwidth}
    \centering
        \includegraphics[width=0.7\columnwidth]{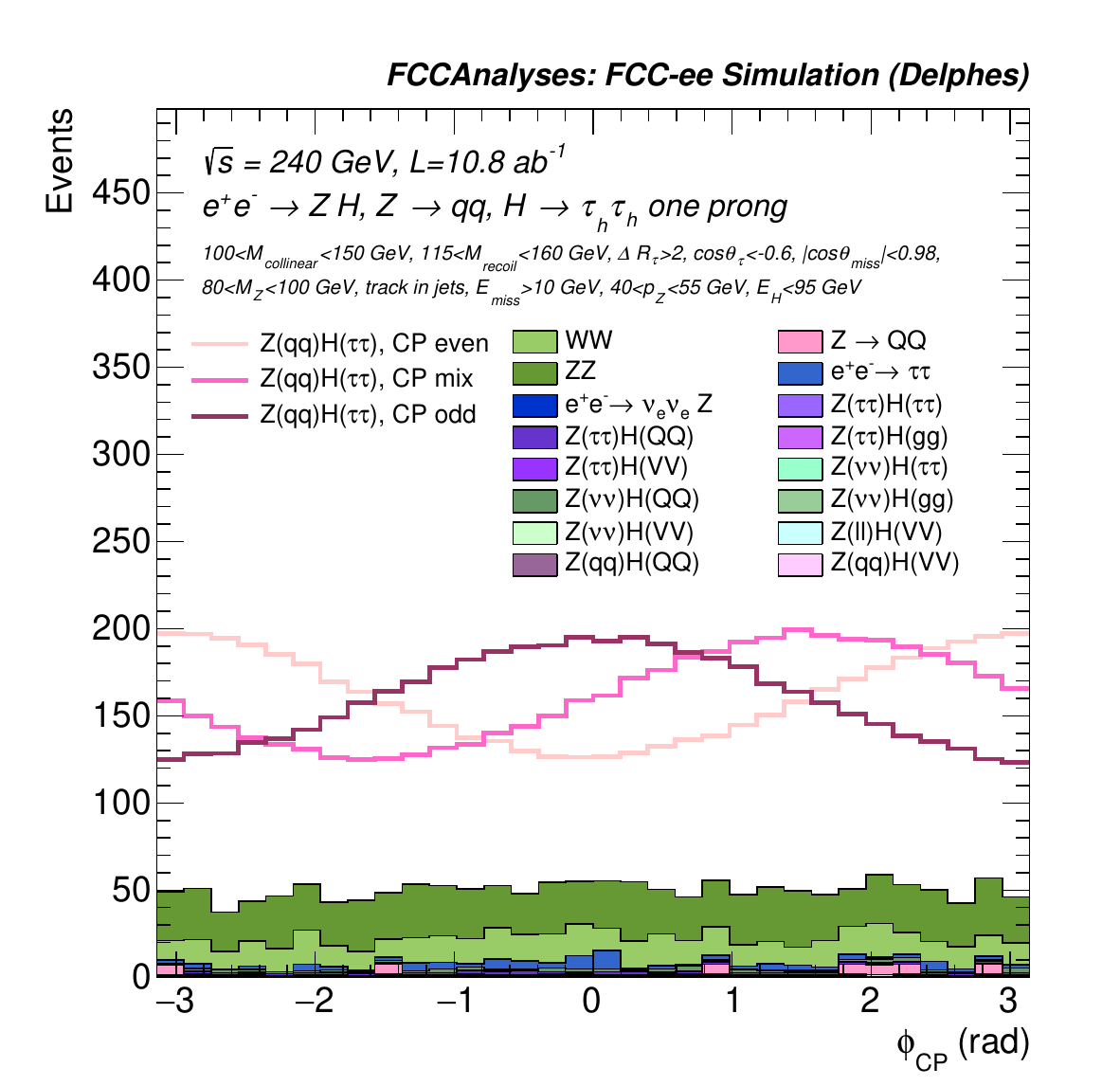}
    \end{subfigure}
    \caption{Distributions for the anomalous coupling signal and backgrounds in the CP-sensitive angular variable. On the top figure: events with Z$\to\ell\ell$. On the bottom figure: events with Z$\to qq$. In both plots, only one-prong hadronic $\tau$ decays are selected. The full kinematic event selection for the respective channels is applied.}
    \label{fig:CP_AC}
\end{figure}

\begin{figure}[htb]
    \centering
    \includegraphics[width=0.7\columnwidth]{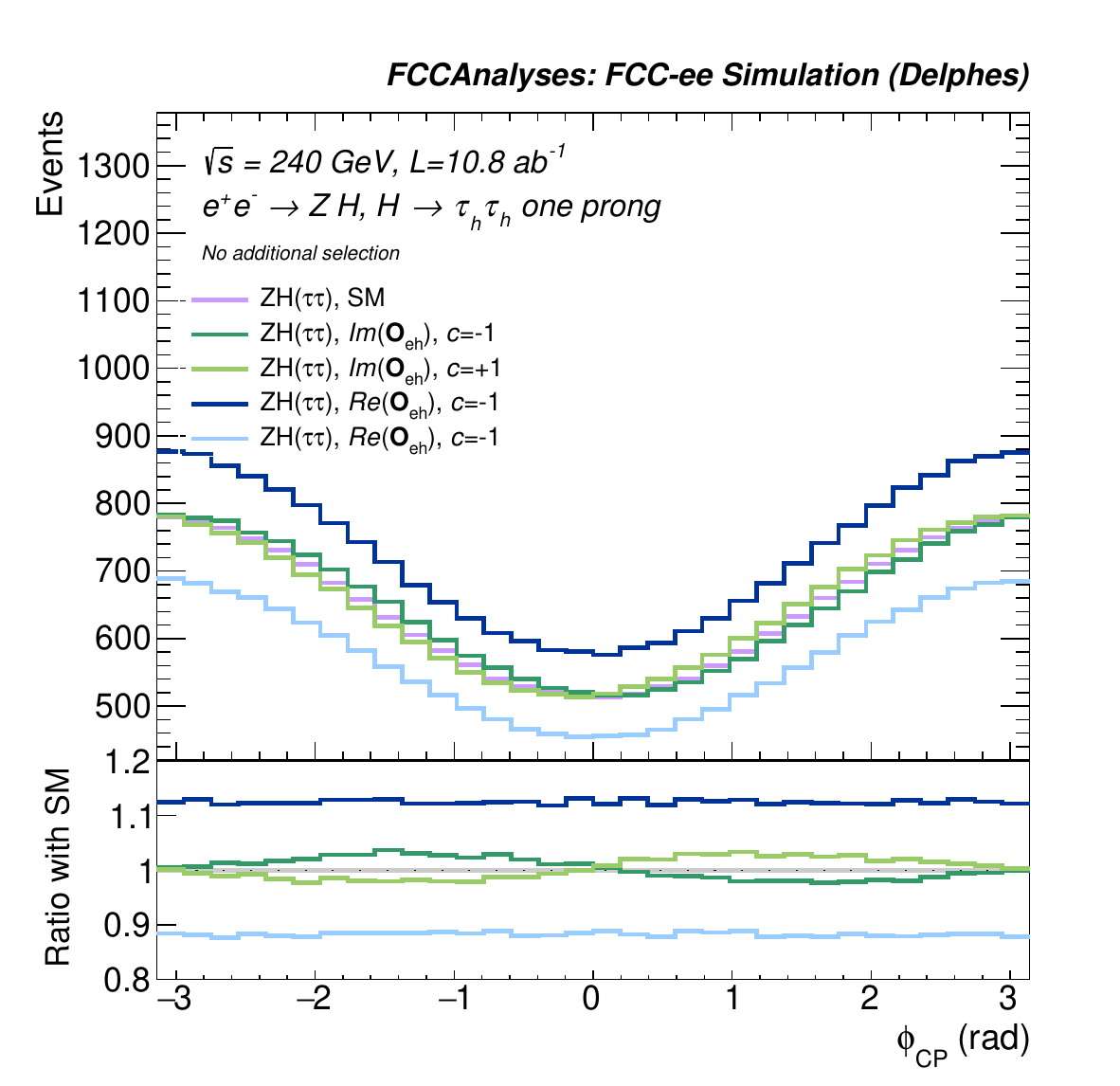}
    \caption{Signal distributions for the SMEFT Yukawa operators in the CP-sensitive angular variable. Only one-prong hadronic $\tau$ decays are selected, while both Z decay channels are combined. Only the object definition and jet clustering are applied here.}
    \label{fig:CP_EFT}
\end{figure}

Figures \ref{fig:CP_AC} and \ref{fig:CP_EFT} illustrate the distributions of signal and background events in $\phi_{CP}$. In the first, it is clearly visible that the modulation of the variable in the signal and the phase shift are introduced by different CP hypotheses. In figure \ref{fig:CP_EFT}, the CP-odd Yukawa operator, proportional to $\mathcal{I}m(C_{eH})$, induces a distinctive phase shift, while the CP-even operator, proportional to $\mathcal{R}e(C_{eH})$, primarily results in a normalization shift of the distribution with minimal shape distortion.

\bigskip
\textit{Results.} We use the CMS Combine tool \cite{combine} to estimate the precision in the determination of the CP-sensitive observable of both signal hypotheses. We consider a log-normal uncertainty of 1\% on the cross-section of the background processes $e^+e^-\to ZZ$ and $e^+e^-\to WW$, Drell-Yan, ZH, and photon-induced processes. No statistical uncertainty from the limited number of simulated events is taken into account, as it does not significantly affect the result.

In the case of the anomalous coupling, we construct a test statistic, based on the one presented in Ref. \cite{CMS:2021sdq}, which uses the signal templates to extract a generic CP mixing angle $\psi_{CP}$:
\begin{align}
    S(&\mu,\psi_{CP})= \\ \notag 
    &\mu \cdot \mathcal{L}_{int}\cdot[ (\cos^2(\psi_{CP}) - \sin(\psi_{CP})\cos(\psi_{CP}))\times\text{EVEN} \\ \notag
    &+ (\sin^2(\psi_{CP}) - \sin(\psi_{CP})\cos(\psi_{CP}))\times\text{ODD} \\ \notag
    &+2\sin(\psi_{CP})\cos(\psi_{CP})\times\text{MIX} ], 
\end{align}
where $\mu$ is the signal strength and 
\begin{gather}
    \text{EVEN} = d\sigma_{\phi_{\tau\tau}=0^\circ}/d\phi_{\tau\tau}, \\ \notag
    \text{MIX} = d\sigma_{\phi_{\tau\tau}=45^\circ}/d\phi_{\tau\tau}, \\ \notag
    \text{ODD} = d\sigma_{\phi_{\tau\tau}=90^\circ}/d\phi_{\tau\tau}, 
\end{gather}
which produces a scan on all angular values. This gives us an estimate on the precision with which the $\phi_{\tau\tau}$ angle could be measured at FCC-ee at 68\% CL, namely $\Delta\phi_{\tau\tau}=\pm2.49^\circ$ when combining all the signal categories, with a further separation of one-prong $\tau_h$ decays. The precision obtained using only this channel is instead of $\Delta\phi_{CP}=\pm2.8^\circ$, demonstrating how these decays carry most of the information needed to distinguish different CP states. This result highlights the potential of the FCC-ee program in constraining the CP state of the Higgs boson, with an improvement of at least a factor of two compared to other future collider options such as HL-LHC and ILC.

In the SMEFT interpretation, the sensitivity of the analysis to dimension-six operators is derived from a likelihood-based statistical model in which the expected event yields depend on the Wilson coefficient vector $\mathbf{c}$.
The predicted number of events in a given phase-space region scales with the Wilson coefficients according to
\begin{equation}
N = N_{\mathrm{SM}} + \sum_\alpha \left[\frac{c_\alpha}{\Lambda^2} N_\alpha^{\text{int}} + \frac{c_\alpha^2}{\Lambda^4} N_\alpha^{\text{quad}}\right] + \sum_{\alpha \neq \beta} \frac{c_\alpha c_\beta}{\Lambda^4} N_{\alpha,\beta}^{\text{mix}},
\label{eq:N_expansion}
\end{equation}
where $N_{\mathrm{SM}}$ accounts for the expected SM process events, and the remaining terms represent linear, quadratic, and interference contributions from the EFT operators, with their magnitudes encoded in the operator-dependent coefficients $N_\alpha^{\text{int}}$, $N_\alpha^{\text{quad}}$, and $N_{\alpha,\beta}^{\text{mix}}$.
The likelihood function $\mathcal{L}$ is constructed to determine the mean values of the Poisson events, $N_k(\mathbf{c})$, with $k$ the bin index of the considered observable. The likelihood is defined as
\begin{equation}
\mathcal{L}(\mathbf{c}) = \prod_k \frac{\left[N_k(\mathbf{c})\right]^{n_k}}{n_k!} e^{-N_k(\mathbf{c})},
\label{eq:SMEFTlikelihood}
\end{equation}
where $\mathbf{c}$ denotes the set of free Wilson coefficients entering the fit, $N_k(\mathbf{c})$ is the expected number of events in bin $k$ following Eq.~(\ref{eq:N_expansion}), and $n_k \equiv N_k(\mathbf{0})$ represents the corresponding SM expectation. 
The expected sensitivity to the Wilson coefficients is estimated based on the likelihood profile by considering different scenarios: one operator left floating at a time, and two operators. In all cases, we estimated both the sensitivity due to linear-only and linear plus quadratic amplitude.

\begin{figure}[htb]
    \centering
    \includegraphics[width=0.8\columnwidth]{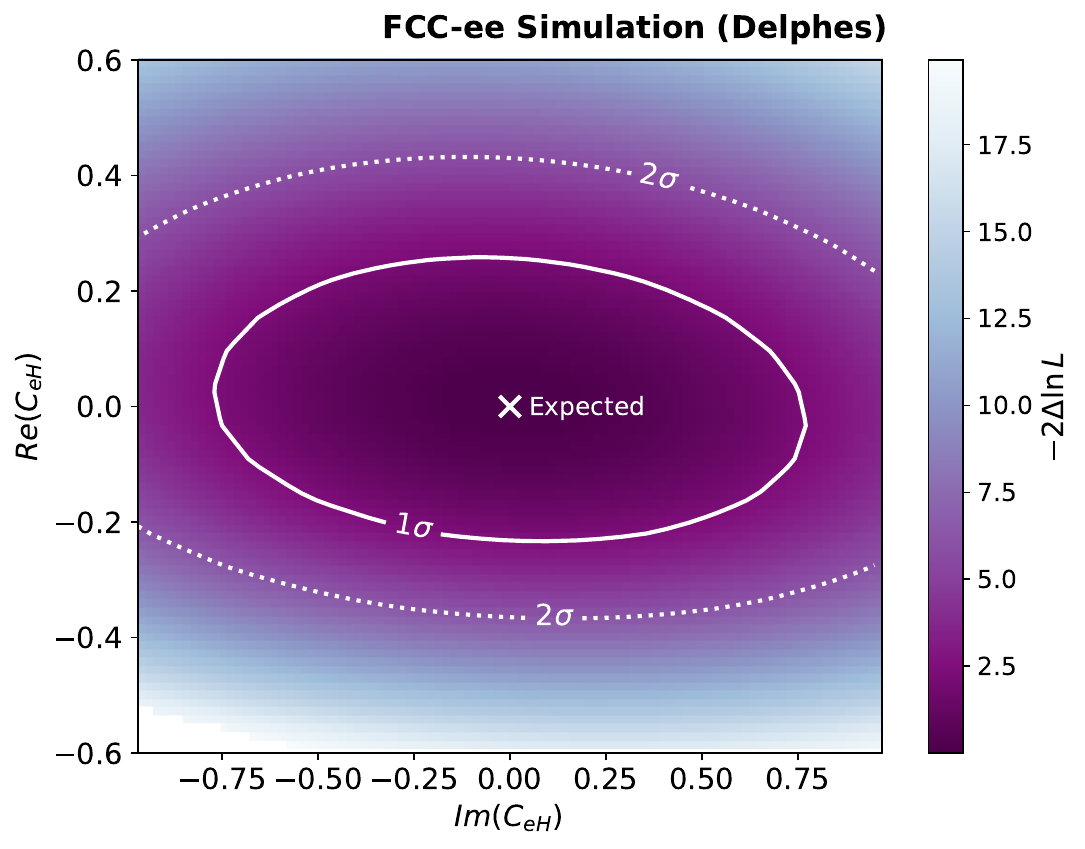}
    \caption{Two-dimensional scan of linear operators $Re(C_{eH})$ and $Im(C_{eH})$, which are the most influential on the CP-odd vertex. The plot shows little correlation between the two.}
    \label{fig:EFT_2d}
\end{figure}

The resulting constraints on these operators are summarized in Figure \ref{fig:EFT_2d}, which displays the two-dimensional likelihood profile in the plane spanned by the real and imaginary parts of the Wilson coefficient $c_{eH}$. The contours are approximately elliptical and aligned with the axes, pointing to a negligible correlation between the CP-even and CP-odd components. The bound on the real part is more stringent than that on the imaginary part, since it is primarily controlled by the overall rate measurement, which is statistically more precise than the shape-driven CP-sensitive observables.
Electric and magnetic dipole moment (EDM and MDM) measurements provide a highly complementary handle on the same operator through Barr–Zee-type two-loop contributions~\cite{Barr:1990vd}. In particular, Ref.~\cite{Kosnik:2025srw} shows that EDM and MDM data impose very strong constraints in the $\mathcal{I}m(\mathcal{O}_{eH})$ direction, at the price of a two-fold degeneracy in the global fit, of which only the solution closer to the SM is phenomenologically favored. In this context, the bounds obtained here play a crucial role: when combined with EDM-based limits, they help resolve the degeneracy and underline the fundamental impact of direct collider measurements, and in particular of the FCC-ee program, in mapping the CP structure of the Higgs–lepton Yukawa interactions.

\begin{figure}[htb]
    \centering
    \includegraphics[width=\columnwidth]{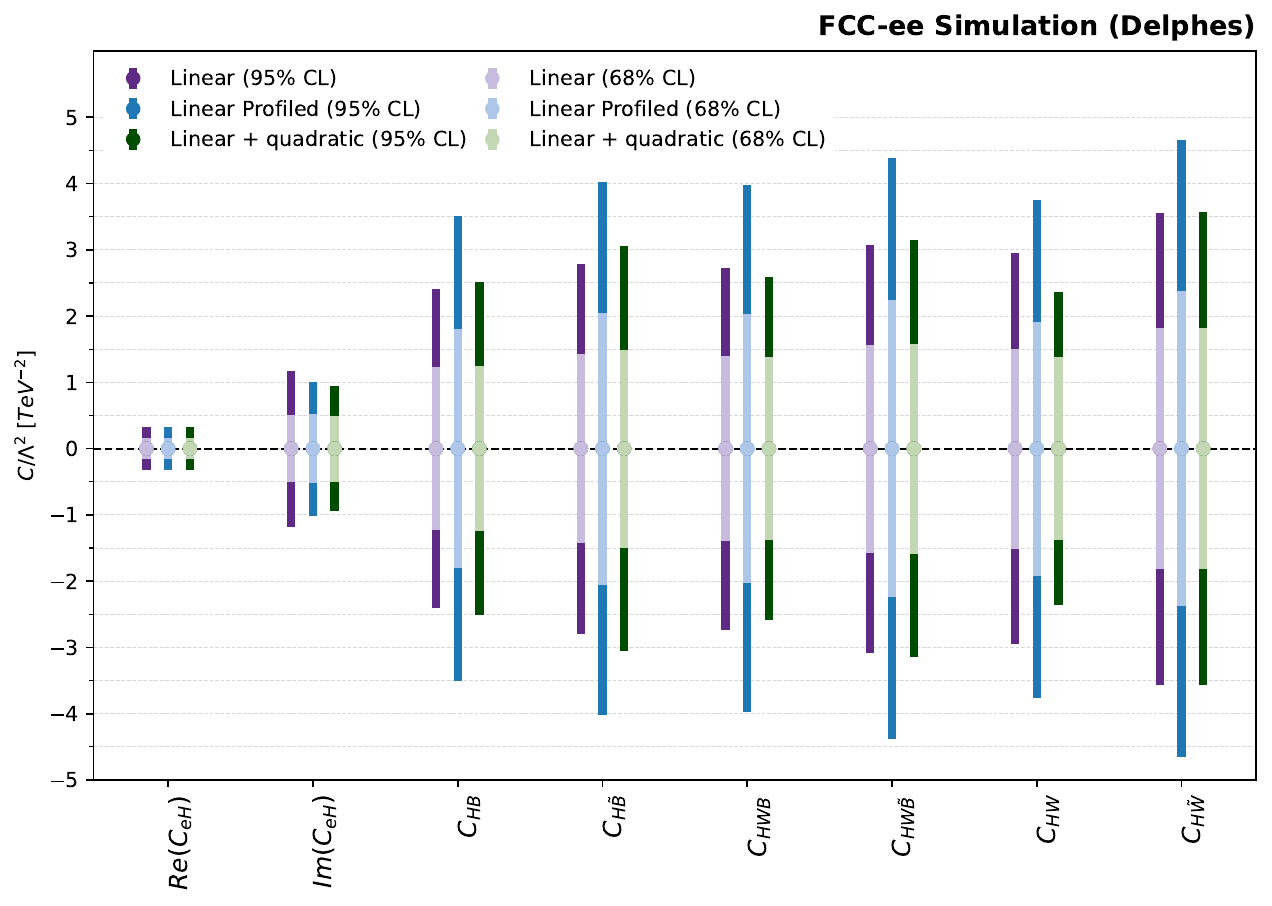}
    \caption{Summary of one-dimensional limits on both linear, profiled linear, and linear plus quadratic EFT operators that act on the H$\to\tau\tau$ vertex.}
    \label{fig:EFT_1d}
\end{figure}

Finally, Figure \ref{fig:EFT_1d} places these results in a broader context by comparing the sensitivity of the Yukawa operators to the CP-odd and CP-even operators affecting the Higgs-gauge boson couplings ($\mathcal{O}_{HW}$, $\mathcal{O}_{HB}$, $\mathcal{O}_{HWB}$). The limits on $\mathcal{R}e(\mathcal{O}_{eH})$ and $\mathcal{I}m(\mathcal{O}_{eH})$ are significantly more stringent, by roughly an order of magnitude, than those obtained for the bosonic operators. This confirms that the $H \to \tau\tau$ channel is a uniquely sensitive probe for CPV in the fermionic sector, while having limited sensitivity to CP effects originating in the gauge-Higgs sector. 
The small degradation of the profiled constraints when fitting all operators simultaneously, relative to single-operator fits, indicates that correlations between CP-even and CP-odd operator directions are limited for the observables considered, leading to only a minor loss of sensitivity.

\bigskip
\textit{Summary.} The FCC-ee provides an exceptionally clean environment to study the CP properties of the Higgs boson through the 
$H\to\tau\tau$ decay, exploiting its large branching fraction, excellent reconstruction, and high-rate ZH production at $\sqrt{s}=$240 GeV. In this work, the CP of the Higgs boson is studied with a fast simulation of the IDEA detector concept, including all relevant background events and Z decays into visible particles, with special attention to the $\tau$ lepton reconstruction. The CP-mixing angle is first constrained in the anomalous coupling framework, achieving an expected precision of $\Delta\phi_{\tau\tau} = \pm 2.5 ^\circ$ at 68\% CL, with most sensitivity arising from one-prong hadronic 
$\tau$ decays, significantly improving on current measurements at the LHC and projections for HL-LHC and ILC. The analysis is then interpreted in the SMEFT framework, where a likelihood built from binned event yields is used to extract bounds on the relevant dimension-six Wilson coefficients in both linear and linear-plus-quadratic truncations, considering one- and two-operator scenarios. The resulting 2D likelihood scan in the $\mathcal{R}e(\mathcal{O}_{eH})$, $\mathcal{I}m(\mathcal{O}_{eH})$ plane shows strong constraints on the CP-even component, driven by the inclusive rate, and only mild correlation between the CP-even and CP-odd directions. When combined with EDM and MDM measurements, which dominantly restrict 
$\mathcal{I}m(\mathcal{O}_{eH})$ but suffer from a two-fold degeneracy, the FCC-ee projections highlight the pivotal role of direct collider measurements in resolving this ambiguity and in mapping the CP structure of the Higgs–lepton Yukawa sector. Finally, one-dimensional limits demonstrate that the leptonic Yukawa operators are constrained roughly an order of magnitude more strongly than CP-odd and CP-even bosonic operators, confirming the $H \to \tau \tau$ channel as a uniquely sensitive probe of fermionic CPV, with limited impact from gauge–Higgs operators.

\bigskip
\begin{acknowledgments}
\textit{Acknowledgments.} This work is supported by the Alexander von Humboldt-Stiftung. The authors are grateful to I. Brivio and A. Cardini for insightful discussions and exchanges.
\end{acknowledgments}

\bibliography{biblio}% Produces the bibliography via BibTeX.

\end{document}